\documentclass[pdflatex,sn-vancouver-ay]{sn-jnl}

\usepackage{graphicx}%
\usepackage{multirow}%
\usepackage{amsmath,amssymb,amsfonts}%
\usepackage{amsthm}%
\usepackage{mathrsfs}%
\usepackage[title]{appendix}%
\usepackage{xcolor}%
\usepackage{textcomp}%
\usepackage{manyfoot}%
\usepackage{booktabs}%
\usepackage{algorithm}%
\usepackage{algorithmicx}%
\usepackage{algpseudocode}%
\usepackage{listings}%
\usepackage{color}
\usepackage{multirow,bigdelim}
\usepackage{rotating}
\usepackage[table]{xcolor}

\newcommand{\E}{\mathbb{E}}
\newcommand{\bias}{\mbox{bias}}
\newcommand{\cov}{{\mbox{cov}}}

\newcommand{\z}{{\bf z}}

\newcommand{\bv}{{\bf v}}

\newcommand{\w}{{\bf w}}
\newcommand{\0}{{\bf 0}}

\newcommand{\X}{{\bf X}}

\newcommand{\y}{{\bf y}}

\newcommand{\bxi}{{\boldsymbol \xi}}
\newcommand{\boeta}{{\boldsymbol \eta}}
\newcommand{\bdelta}{{\boldsymbol \delta}}

\newcommand{\bbeta}{{\boldsymbol \beta}}
\newcommand{\bvarepsilon}{{\boldsymbol \varepsilon}}

\newcommand{\bu}{{\bf u}}
\newcommand\ind{\bot\hspace*{-6pt}\bot}  

\theoremstyle{thmstyleone}%
\newtheorem{theorem}{Theorem}
\theoremstyle{thmstyletwo}%

\theoremstyle{thmstylethree}%
\newtheorem{lemma}{Lemma}%

\begin{document}
\newgeometry{margin=3cm}

\title[Uncertainty intervals for multilevel models with missing not at random data]{Uncertainty intervals for multilevel models with missing not at random data}

\author[]{\fnm{Minna}\sur{Genb\"ack}}\email{minna.genback@umu.se}

\affil[]{\orgdiv{Ume\r{a} School of Business Economics and Statistics}, \orgname{Ume\r{a} University}, \orgaddress{901 87 Ume\r{a}, Sweden}}

\abstract{We propose a sensitivity analysis method for missing not at random (MNAR) data in the context of linear multilevel (mixed-effects) models. The outcome and dropout risk are both modelled using multilevel models and a bias adjustment due to MNAR data is derived. This bias can be estimated from observed data conditional on specified values of sensitivity parameter(s). Under the assumption that these parameters lie within a plausible range, the method partially identify the parameters of interest, yielding bounds for estimation and inference under assumptions weaker than missing at random. The proposed analysis is investigated in a simulation study and illustrated with an analysis of the association between loneliness and physical activity with memory trajectories, adjusting for demographic, socioeconomic, and health covariates.}


\maketitle

\section*{Acknowledgements}
The work was funded by Swedish Research Council for health working life and welfare [2019-01064]. The author thank associate professors Maria Josefsson, Anders Lundquist and Xavier de Luna for helpful discussions. This paper uses data from SHARE wave 4-9. The SHARE data collection has been primarily funded by the European Commission, see www.share-eric.eu for more details.

\section{Introduction}
\label{s:intro}

In longitudinal surveys,  non-response is a common occurrence and it comprises not only of individuals who never participate but also those who respond at only a subset of the measurement occasions. Non-response can reduce study quality—firstly by reducing sample sizes and thus the precision of estimates, and more severely, by introducing bias into the estimation of the parameters of interest, if certain assumptions about the missing data are not met. Analyses of data with non-response necessarily rely on untestable assumptions about the missingness mechanism to achieve full identification of the parameters of interest \citep{molenberghs2015}.

A widely used method for the analysis of longitudinal data is the multilevel (mixed-effects) models. Standard implementations of these models assume that data are missing at random (MAR), that is the probability of missingness only depend on the observed data \cite{LR:02}. Among older populations,  previous studies suggest that respondents tend to be healthier than non-respondents \citep{chatfield2005, josefsson2012, nummela2011}. If so, in studies where the outcome is health-related, the MAR assumption may not fulfilled and therefore the conclusions if incorrectly assuming MAR could be misleading. 

Most methods that accommodate MNAR mechanisms fall into either selection models or pattern mixture models. In selection models, dropout is modeled as a dependent variable, whereas in pattern mixture models, it is treated as a predictor \citep{enders2023, carpenter2013}. Within both frameworks, there are approaches that achieve full identification under alternative assumptions to MAR, such as additional distributional constraints \citep{enders2011, josefsson2016, hammon2020}. These assumptions are untestable from the data, and if made incorrectly may lead to misleading conclusions.

An alternative approach, proposed in this paper, is to forgo full identification and instead adopt weaker assumptions about the missing data mechanism, thereby partially identifying the parameters of interest \citep{manski2003, vansteelandt2006}.  This strategy typically involves quantifying deviations from the MAR assumption using sensitivity parameter(s), where the width and direction of the resulting bounds reflect the extent to which the sensitivity parameter departs from MAR. In this paper, we implement this by fitting two multilevel models: one for the outcome under the MAR framework and another for the dropout mechanism. Under MAR, the random components of these two regression models are independent; here, we allow them to be correlated, with correlation serving as the sensitivity parameter. This type of sensitivity parameter has previously been applied in settings other than multilevel models \citep{genback2015, genback2019, Imai:2010}. We provide a general analytical expression for the bias as a function of the sensitivity parameter, which, to the authors’ knowledge, has previously only been derived for special cases that are not representative of longitudinal studies \citep{G:04, GF:04}. The bias is then used to construct uncertainty intervals, analogous to confidence intervals but centered around bounds on the estimates, thereby allowing greater flexibility in assumptions about the missing data mechanism. This type of analysis is to the authors knowledge not common for multilevel models but have recently been suggested in a  multiple imputation setting \cite{Wang2025}. 


We apply the proposed methods to a study of the association between various lifestyle factors and memory trajectories among individuals aged 65 and older, using multilevel models fitted to longitudinal survey data. The survey data exhibit substantial dropout, which we suspect may follow a MNAR mechanism. In such cases, standard multilevel analyses may yield biased estimates. To address this, we adjust the standard analysis to account for uncertainty arising from the missing data.

The paper is organized as follows. Section~\ref{Sec.Th} introduces the general notation, model assumptions, and theoretical results. In Section~\ref{Sec.ex}, we describe the motivating example and specialize the notation and assumptions to this setting. Section~\ref{Results} presents the results of the data analysis, and Section~\ref{Simulation} describes a simulation study designed to assess the performance of the proposed sensitivity analysis. The paper is concluded in Section~\ref{Conclusion}.

\section{Theory}
\label{Sec.Th}
 \noindent  We model the outcome with a multilevel model (linear mixed effects model). Specifically, suppose that we have $N$ clusters, each with $n_{i}$ observations. Let $\y_{i}$ be an $n_{i} \times 1$ vector containing the outcomes, $\X_{i}$  a $n_{i}\times m$ matrix of covariates and $\bbeta$ an $m \times 1$ vector of the parameters of interest. Then,
  \begin{align}
 \y_{i} = \X_{i}\bbeta +  \boeta_{i} ,
\label{out.eq}
\end{align}
where $  \boeta_{i} $ is multivariate normal with covariance matrix $\Sigma_{i}$, thereby allowing for dependencies between the observations within each cluster. Note that, to ease notation we only use 2 levels in this section, but the results are easily adaptable to more levels by introducing more indexes, in particular we use 3 levels in the application in Section \ref{Sec.ex}.

Dropout is modeled with a generalized multilevel model (generalized linear mixed model) with a probit link. That is, let $\z_{i}^*$ be a latent $n_{i} \times 1$ vector with elements $z_{ik}^*$, $k=1,...,n_{i}$ where $y_{ik}$ is observed if $z_{ik}=I(z_{ik}^*>0)=1$, and let $\bdelta$ be an $m \times 1$ parameter vector. Then,

 \begin{align}
 \z_{i}^* =  \X_{i}\delta +   \bu_{i},
\label{sel.eq}
\end{align}
where $  \bu_{i} $ is multivariate normal with covariance matrix $\Omega_{i}$, again allowing for within cluster dependence.

 Most statistical softwares that can fit both models (\ref{out.eq}) and (\ref{sel.eq}) under the assumption that dropout is MAR, i.e. $\bu_i \ind \boeta_i$.  To allow for MNAR data, we introduce a sensitivity parameter matrix, $\Delta_i$, the covariance between $\bu_i$ and $\boeta_i$. We can then jointly specify the outcome and dropout processes as:
 \begin{align}
\left[ \y_{i} \atop \z_{i}^* \right] = \left[ \X_{i}\bbeta \atop \X_{i}\delta \right] +  \left[ \boeta_{i} \atop \bu_{i} \right], \mbox{ where }  \left[ \boeta_{i} \atop \bu_{i} \right] \sim MVN\left(\left[ \0 \atop \0 \right], \left[{ \Sigma_{i} \; \Delta_{i}} \atop { \Delta_{i}^T \; \Omega_{i}} \right]   \right).
\label{joint.eq}
\end{align}


\subsection{Estimator and bias}
\noindent  Let $\y_{i}^{\text{obs}}$ denote the observed subset of the outcome vector for cluster $i$ ($\y_{i}^{\text{obs}}=\{ y_{ik} | z_{ik}=1, k=1,...,n_{i}$\} ) and $\{\}^{\text{obs}}$  denote the corresponding subset for any general vector.  Similarly, let $\X_i^{{\text{obs}}}$ be the observed subset of the covariate matrix, and $\Sigma_i^{{\text{obs}}}$ the corresponding submatrix of $\Sigma_i$. Then, the generalized least squares estimator (GLS) of $\bbeta$ under model (\ref{out.eq}) using the observed data is \citep{Demidenko:2004}, 

\begin{equation}
\tilde \bbeta_{GLS} =\left[\sum_{i=1}^N (\X_{i}^{\text{obs}})' (\Sigma_{i}^{\text{obs}})^{-1}  \X_{i}^{\text{obs}} \right]^{-1} \left[\sum_{i=1}^N  (\X_{i}^{\text{obs}})' (\Sigma_{i}^{\text{obs}})^{-1}  \y_{i}^{\text{obs}} \right].
\label{GLS}
\end{equation}
As previously noted, this estimator may be biased if dropout is MNAR.

\begin{theorem}
\label{proof.bias}
The bias of the the generalized least squares estimator (\ref{GLS}) of $\bbeta$ under the model (\ref{joint.eq}), when $\Delta_{i} \neq \0$, is:

\begin{align*}
\bias(\tilde \bbeta_{GLS} ) = \left[\sum_{i=1}^N (\X_{i}^{\text{obs}})' (\Sigma_{i}^{\text{obs}})^{-1}  \X_{i}^{\text{obs}} \right]^{-1}\left[\sum_{i=1}^N  (\X_{i}^{\text{obs}})' (\Sigma_{i}^{\text{obs}})^{-1} \left\{  \Delta_{i} \Omega_{i}^{-1}   E\left( \bv_i  \right)\right\}^{\text{obs}} \right]
\end{align*}

where $\bv_i=\left\{\bu_i \left| \left[ \bu_i^{\text{obs}} > -{\X_i}^{\text{obs}} \delta \atop \bu_i^{mis} < -{\X_i}^{mis} \delta\right], \X_i \right.\right\}$ is a double truncated multivariate normal distribution with mean zero and covariance matrix $\Omega_i$. The truncations are from below if $z_{ik}=1$, and from above if $z_{ik}=0$, both at $-{\bf{x}}_{ik}\delta$. 
\begin{proof}

\begin{align*}
&\bias(\tilde \bbeta_{GLS} )= \E\left( \tilde \bbeta_{GLS}  \right)-\bbeta =\E\left(\left[\sum_{i=1}^N (\X_{i}^{\text{obs}})'  (\Sigma_{i}^{\text{obs}})^{-1} \X_{i}^{\text{obs}} \right]^{-1} \left[\sum_{i=1}^N (\X_{i}^{\text{obs}})'  (\Sigma_{i}^{\text{obs}})^{-1} \y_{i}^{\text{obs}} \right] \right) -\bbeta\\
&=\E\left(\left. \E\left(\left[\sum_{i=1}^N (\X_{i}^{\text{obs}})'  (\Sigma_{i}^{\text{obs}})^{-1} \X_{i}^{\text{obs}} \right]^{-1} \left[\sum_{i=1}^N (\X_{i}^{\text{obs}})'  (\Sigma_{i}^{\text{obs}})^{-1} \y_{i}^{\text{obs}} \right] \right| \X \right) \right) -\bbeta\\
&=\E \left( \left[\sum_{i=1}^N (\X_{i}^{\text{obs}})'  (\Sigma_{i}^{\text{obs}})^{-1} \X_{i}^{\text{obs}} \right]^{-1} \left[\sum_{i=1}^N (\X_{i}^{\text{obs}})'  (\Sigma_{i}^{\text{obs}})^{-1}\E\left(\left.  \y_{i}^{\text{obs}}\right|\X\right)  \right] \right) -\bbeta\\
&=\E \left( \left[\sum_{i=1}^N (\X_{i}^{\text{obs}})'  (\Sigma_{i}^{\text{obs}})^{-1} \X_{i}^{\text{obs}} \right]^{-1} \left[\sum_{i=1}^N (\X_{i}^{\text{obs}})'  (\Sigma_{i}^{\text{obs}})^{-1}\left( \X_{i}^{\text{obs}} \bbeta  +\left\{  \Delta_{i} \Omega_{i}^{-1}   E\left( \bv_i\right)\right\}^{\text{obs}} \right)  \right] \right) -\bbeta\\
&=  \E \left( \left[\sum_{i=1}^N (\X_{i}^{\text{obs}})'  (\Sigma_{i}^{\text{obs}})^{-1} \X_{i}^{\text{obs}} \right]^{-1} \left[\sum_{i=1}^N (\X_{i}^{\text{obs}})'   (\Sigma_{i}^{\text{obs}})^{-1}\left\{  \Delta_{i} \Omega_{i}^{-1}   E\left(\bv_i \right)\right\}^{\text{obs}} \right] \right),
\end{align*}
where the second to last last equality follows from Lemma \ref{proof.exp} in Appendix \ref{appTh}.
\end{proof}
\end{theorem}

It is possible to estimate $\Omega_i^{-1}$ and $\bdelta$ from model (\ref{sel.eq}), and $\Sigma_i^{\text{\text{obs}}}$ from model (\ref{out.eq}) using most statistical software, such as {\tt lme4} in {\tt R} \citep{lme4, R}. However, convergence issues may arise when using complex covariance structures. We estimate $\mathbb{E}(\bv_i)$ using the {\tt R} package {\tt MomTrunc} \citep{MomTrunc}, plugging in the estimate of $\bdelta$. Thus, for fixed values of $\Delta_i$, Theorem \ref{proof.bias} enable us to consistently estimate the bias due to MNAR dropout. 

Note that, similar sensitivity analyses have been proposed previously \citep{G:04, GF:04}, though without applications or simulations. These works suggest that $\tilde \bbeta_{\text{GLS}}$ follows a closed skew-normal or SUN distribution, which holds only under the, for longitudinal settings, unrealistic assumption that each cluster has either all outcomes observed or all missing. Grilli and Rampichini \cite{GR:10} discuss bias in special cases of the model but do not provide analytical expressions for the bias, as we do here.

\subsection{Sensitivity analysis}
\label{Sens.sec}
As mentioned in the previous section, for fixed values of the sensitivity parameter matrix $\Delta_i$, the bias of this estimator can be quantified using Theorem \ref{proof.bias}, under the modelling assumptions stated at the beginning of this section. Thus, point estimates and confidence intervals for $\bbeta$ can be constructed by combining the two—provided that $\Delta_i$ is known, i.e., under specific MNAR assumptions.

While $\Delta_i$ is general in the sense that it accommodates a wide range of missingness mechanisms, it may be difficult to specify and interpret in practice. We therefore recommend defining one or more univariate sensitivity parameters to clarify the assumptions about the missing data mechanism, and using these to construct $\Delta_i$ (see the example in Section \ref{Sec.ex}).

Even with univariate sensitivity parameters, it is rarely realistic to know their exact values. We therefore suggest specifying a plausible range for each sensitivity parameter, which yields a corresponding range of point estimates and confidence intervals for $\bbeta$. If the true values of the sensitivity parameters lie within these pre-specified ranges, then the union of the resulting confidence intervals—referred to as uncertainty interval—will cover $\bbeta$ with at least $(1 - \alpha)\%$ probability. Uncertainty intervals is a useful compliment to the MAR analysis to illustrate the stability of the conclusions to different assumptions. For a more detailed theoretical discussion of uncertainty intervals \citep{vansteelandt2006, genback2019}.

\section{Motivating example}
\label{Sec.ex}
\subsection{Data}
\noindent We use data from the Survey of Health, Ageing and Retirement in Europe (SHARE) \citep{share_w4, share_w5, share_w6, share_w7, share_w8, share_w9}. SHARE is a cross-national longitudinal survey conducted approximately every two years across various European countries \cite{share_dataset}. Data are collected through computer-assisted personal interviews and include extensive information on objective and subjective health measures, socioeconomic status, and social networks. The sample is defined on household level, meaning that all individuals in sampled households are included.

We use data from waves 4–9, collected between 2011 and 2022. Wave 4 is chosen as the baseline, as a large refreshment sample was introduced in this wave. At baseline, 29,613 individuals aged 65 or older participated, of whom 2,300 had partially missing data on one or more variables of interest and were therefore excluded. This results in a final study sample of 27,313 individuals across 20,981 households. In each wave, efforts are made to re-interview all sampled individuals. However, a substantial proportion of the baseline sample either drop out entirely or miss one or more waves. Some of the missing data are due to countries opting not to participate in certain waves; these are assumed to be missing at random. In waves 5 through 9, respectively, $25\%$, $37\%$, $46\%$, $68\%$, and $67\%$ of the study sample in participating countries are missing. It is important to note that death is poorly recorded in SHARE, making it difficult to distinguish between attrition and death for most individuals; this is therefore not addressed in the analysis.

Our aim is to investigate the association between baseline loneliness and physical activity with memory trajectories, adjusting for demographic, socioeconomic, and health-related covariates, for individuals aged 65 and above. The outcome of interest is a memory score, based on the number of words (out of 10) recalled immediately and after a delay, yielding a total score between 0 and 20.

Loneliness is measured using the question: “How often do you feel left out of things?” with response options: often, sometimes, rarely, or never. Physical activity is assessed by the question: “How often do you engage in vigorous physical activities such as sports, heavy housework, or a job that involves physical labor?” and is categorized as $\geq 1$/week versus $1–3$/month or less.
 
 Control variables are collected at baseline and include age at baseline (in 2011), region (north: Sweden, Denmark, Estonia; west: France, Austria, Germany, Netherlands, Switzerland, Belgium; east: Czech Republic, Slovenia, Hungary, Poland; south: Spain, Italy, Portugal), living alone, education level (based on ISCED 1997, categorized as $\leq 2$, $3-4$ or $\geq 5$), self-rated health (recoded into very good: excellent or very good; good: good; or poor: fair or poor), depression ("In the last month, have you been sad or depressed?") and time (years since baseline).
 
\subsection{Notation and setting, example}
 \label{seq.model.ex}
 \noindent Suppose that we have $N$ households, with $H_i$ individuals in household $i$, and $n_{ij}$ observations for each individual $j=1, ... , H_i$ in household $i=1,...,N$. 

\vspace{1em}
\noindent\textbf{Outcome model}
\vspace{0.5em}

 A special case of the outcome regression model (\ref{out.eq}) can be rewritten as:
\begin{equation}
 \y_{ij}=\X_{ij} \bbeta +(a_i + b_{ij}) +(c_{i} + d_{ij}) \mbox{\bf{time}}_{ij} +\bvarepsilon_{ij},\;\; i=1,\;...,\;N,\;\; j=1,\;...,\;H_i,
\label{Ex.out.eq}
\end{equation}
where $\bvarepsilon_{ij}$ is an $n_{ij} \times 1$ vector of i.i.d. normally distributed error terms, independent of the random effects  ($a_i$, $b_{ij}$, $c_i$, $d_{ij}$), which are also normally distributed with mean zero. The random effects allow for between-individual variation in baseline memory (random intercept $b_{ij}$) and change over time (random slope $d_{ij}$), as well as between-household variation ($a_i $ and $c_i$).

\vspace{1em}
\noindent\textbf{Dropout model}
\vspace{0.5em}

The sensitivity analysis accounts for dropout but not for initial selection into the study. That is, we condition on participation in wave 4 (baseline). Thus, the generalized multilevel model for dropout includes a fixed intercept. The dropout model (\ref{sel.eq}) under these conditions is: 
\begin{equation}
\z^*_{ij}= \X_{ij} \bdelta +(\nu_i+\gamma_{ij} ) \mbox{\bf{time}}_{ij} +\bxi_{ij},\;\; i=1,\;...,\;N,\;\; j=1,\;...,\;H_i,
\label{Ex.sel.eq}
\end{equation}
where $\z_{ij}^*$ is an $n_{ij} \times 1$ latent vector with elements $z_{ijk}^*$, $k = 1, \dots, n_{ij}$, and $y_{ijk}$ is observed if $z_{ijk} = I(z^{ijk} > 0) = 1$. The error terms $\xi_{ijk}$ are i.i.d. normal with standard deviation $\sigma_\xi$, and the individual-specific random slopes $\gamma_{ij}$ are independent of $\bxi_{ij}$, and normally distributed with mean zero.

\vspace{1em}
\noindent\textbf{Sensitivity  parameter matrix}
\vspace{0.5em}

As described in Section \ref{Sec.Th}, the sensitivity parameter $\Delta_i$ represents the covariance between the random components of the outcome and dropout models. Under MAR, $\Delta_i = \0$, and the linear mixed effects model yields consistent estimates of $\bbeta$. We allow $\Delta_i$ to be an arbitrary covariance matrix, but it must be specified in advance for the sensitivity analysis.
In this example, we hypothesize that individuals with declining memory are more likely to drop out. Thus, we allow dropout risk to be associated with the outcome through the random slopes in equations (\ref{Ex.out.eq}) and (\ref{Ex.sel.eq}). Specifically, we define as sensitivity parameters the following correlations: $\rho_{\text{hh}} = \text{cor}(c_i, \nu_i)$, $\rho_{\text{ind}} = \text{cor}(d_{ij}, \gamma_{ij})$, and $\rho_{\text{err}}=\text{cor}(\bxi_{ij}, \bvarepsilon_{ij})$, assuming that they all are within $[0, 0.5]$, consistent with the idea both long term and momentary memory decline is associated with increased long term and momentary dropout risk. We assume no other marginal associations between the random components of the outcome and dropout models.

Given the nested design, elements of $\Delta_i$ (the sensitivity parameter matrix for household $i$) depend on whether observations come from the same individual ($j = h$) or different individuals ($j \neq h$), and on the time points $k$ and $l$. We use estimates of $\sigma_a, \sigma_b, \sigma_c, \sigma_d, \rho_{ac}$, and $\rho_{bd}$ to derive $\Delta_i$ for different values of $\rho_{\text{hh}}$, $\rho_{\text{ind}}$ and $\rho_{\text{err}}$. 

\begin{align*}
\Delta_{i\{jh\}\{kl\}}&=\mbox{cov}(\eta_{ijk},u_{ihl})=\cov((a_i + b_{ij}) +(c_{i} + d_{ij}) time_{ijk} + \varepsilon_{ijk}, (\nu_i+\gamma_{ih} )  time_{ihl}+ \xi_{ihl})\\
&=...= M_1 +M_2 +M_3
\end{align*}
where
\begin{align*}
M_1&=time_{ihl}(\rho_{ac}\rho_{\text{hh}}\sigma_a\sigma_\nu)+time_{ijk}time_{ihl}(\sigma_c\sigma_\nu\rho_{\text{hh}}) \qquad \qquad\qquad \qquad\qquad\quad \\
M_2&= \left[  time_{ihl}(\rho_{bd}\rho_{\text{ind}}\sigma_b \sigma_\gamma)+time_{ijk}time_{ihl}( \sigma_d\sigma_\gamma \rho_{\text{ind}})  \right] I(j=h) \\
M_3&=[ \rho_{\text{err}} \sigma_\varepsilon \sigma_\xi ] I( j=h \& k=l)
\end{align*}
Note that, we rely on estimates calculated under MAR assumptions to specify $\Delta_i$ thereby risking misspecification of the sensitivity parameter matrix. However, directly specifying $\Delta_i$ is too complex to be a  realistic alternative. Note also that, if we fit an unnested model (i.e. ignore the household clustering) then $M_1=0$, making the specification of $\Delta_i$ less cumbersome.

\subsection{Results}
\label{Results}
There are notable differences in baseline characteristics between the subsample that participated in all waves and the subsample that opted out of at least one wave. The always  participants were, on average, younger, had higher levels of education, reported better self-rated health, were less likely to frequently feel left out, and were more likely to engage in vigorous physical activity at least once per week (see Table \ref{desc.tab}). 

Assuming MAR, frequently feeling left out was associated with an average decrease of 0.59 (95\% CI: 0.47, 0.70) words in memory score (out of 20 words) over time, while engaging in vigorous physical activity at least once per week was associated with an average increase of 0.41 (95\% CI: 0.34, 0.48) words. Sensitivity analyses suggest that if dropout is related to poorer memory, these estimates may be slightly low, but the overall conclusions remain similar (see Table \ref{res.tab}). The multilevel models for both the outcome and dropout processes, equations (\ref{Ex.out.eq})–(\ref{Ex.sel.eq}), were also fitted without accounting for household clustering, yielding nearly identical results (not shown).

\begin{table}[h!]
\small
\centering
\begin{tabular}{llllll}
& Complete & Missing & Total \\ 
&(N=5628) &(N=21685) &(N=27313) \\ 
  \hline
{\bf age in 2011} &  \\ 
\quad mean (sd) & 71.1 (4.9) & 74.8 (6.9) & 74.0 (6.7)   \\ 
{\bf region} &  \\ 
\quad south & 923 (16.4\%) & 3660 (16.9\%) & 4583 (16.8\%) \\ 
\quad east & 1054 (18.7\%) & 4823 (22.2\%) & 5877 (21.5\%)   \\ 
\quad west & 2249 (40.0\%) & 8863 (40.9\%) & 11112 (40.7\%)   \\
\quad north & 1402 (24.9\%) & 4339 (20.0\%) & 5741 (21.0\%) \\ 
\bf education &  \\ 
\quad 0-2 & 2346 (41.7\%) & 11377 (52.5\%) & 13723 (50.2\%) \\ 
\quad 3-4 & 2057 (36.5\%) & 6889 (31.8\%) & 8946 (32.8\%)  \\ 
\quad 5-8 & 1225 (21.8\%) & 3419 (15.8\%) & 4644 (17.0\%)   \\ 
\bf{living alone} &  \\ 
\quad not alone & 4151 (73.8\%) & 15498 (71.5\%) & 19649 (71.9\%)  \\ 
\quad alone & 1477 (26.2\%) & 6187 (28.5\%) & 7664 (28.1\%)   \\ 
\bf{self-rated health} &  \\ 
\quad poor &  2061 (36.6\%) & 11553 (53.3\%) & 13614 (49.8\%)   \\ 
\quad good & 2147 (38.1\%) & 6893 (31.8\%) & 9040 (33.1\%)   \\ 
\quad very good & 1420 (25.2\%) & 3239 (14.9\%) & 4659 (17.1\%)   \\ 
\bf{depressed} &  \\ 
\quad no &  3418 (60.7\%) & 12617 (58.2\%) & 16035 (58.7\%)   \\ 
\quad yes & 2210 (39.3\%) & 9068 (41.8\%) & 11278 (41.3\%)   \\ 
\bf{feel leftout} &  \\ 
\quad not often & 5358 (95.2\%) & 19553 (90.2\%) & 24911 (91.2\%) \\ 
\quad often & 270 (4.8\%) & 2132 (9.8\%) & 2402 (8.8\%)   \\ 
 \bf{vigorous} &  \\ 
\quad not active &  2847 (50.6\%) & 14258 (65.8\%) & 17105 (62.6\%)  \\ 
\quad active &2781 (49.4\%) & 7427 (34.2\%) & 10208 (37.4\%)  \\ 
   \hline
\end{tabular}
 \caption{Baseline characteristics of the subsample that participated in all available waves (complete), the subsample that opted out of at least one wave (missing), and the total study sample. Note that some countries only participated in a subset of waves 5–9; therefore, for these countries, the complete subsample have participated in fewer than six waves. For categorical variables, the number of individuals and the corresponding percentages for each response category are presented.}
 \label{desc.tab}
\end{table}

\begin{table}[h!]
\small
\centering
\begin{tabular}{lllllll}
&  \multicolumn{2}{c}{MAR}& \multicolumn{2}{c}{MNAR {\footnotesize ($\rho_{\text{hh}}$, $\rho_{\text{ind}}$) $\in [0, 0.5]$}} &\multicolumn{2}{c}{MNAR {\footnotesize ($\rho_{\text{hh}}$, $\rho_{\text{ind}}$, $\rho_{\text{err}}$) $\in [0, 0.5]$}}\\
 \cmidrule(lr){2-3} \cmidrule(lr){4-5} \cmidrule(lr){6-7} 
 & est& $95\%$ ci & est& $95\%$ ui & est &$95\%$  ui \\ 
  \hline \smallskip
\bf{(Intercept) }&    5.73 & ( 5.64,  5.83) & ( 5.73,  5.77) & ( 5.64,  5.86) & ( 5.55,  5.77) & ( 5.45,  5.86) \\ 
{\bf age in 2011} & -1.22 & (-1.26, -1.19) & (-1.24, -1.20) & (-1.27, -1.16) & (-1.31, -1.20) & (-1.34, -1.16) \\ 
 {\bf region}  \\
\quad  east &  0.60 & ( 0.49,  0.71) & ( 0.58,  0.61) & ( 0.47,  0.72) & ( 0.56,  0.61) & ( 0.46,  0.72) \\ 
\quad west &   1.31 & ( 1.21,  1.41) & ( 1.28,  1.31) & ( 1.19,  1.41) & ( 1.25,  1.31) & ( 1.15,  1.41) \\ 
 \quad north & 1.08 & ( 0.97,  1.19) & ( 1.07,  1.08) & ( 0.96,  1.19) & ( 1.07,  1.10) & ( 0.96,  1.21) \\ 
  \bf{education} \\
  \quad $3-4$ &1.22 & ( 1.14,  1.29) & ( 1.21,  1.22) & ( 1.14,  1.29) & ( 1.21,  1.23) & ( 1.14,  1.30) \\ 
   \quad $5-8$ &  2.23 & ( 2.14,  2.32) & ( 2.23,  2.24) & ( 2.13,  2.33) & ( 2.23,  2.27) & ( 2.13,  2.37) \\ 
\bf{living alone} &   0.20 & ( 0.13,  0.27) & ( 0.20,  0.25) & ( 0.13,  0.32) & ( 0.20,  0.29) & ( 0.13,  0.37) \\ 
\bf{self-rated health} &  \\ 
\quad good & 0.65 & ( 0.57,  0.72) & ( 0.64,  0.66) & ( 0.56,  0.73) & ( 0.64,  0.71) & ( 0.56,  0.78) \\ 
  \quad     very good &     1.03 & ( 0.93,  1.12) & ( 1.01,  1.04) & ( 0.92,  1.13) & ( 1.01,  1.11) & ( 0.92,  1.21) \\ 
\bf{depressed} &-0.03 & (-0.10,  0.03) & (-0.04, -0.03) & (-0.10,  0.04) & (-0.04, -0.01) & (-0.10,  0.05) \\ 
  \bf{feel left out often}& -0.59 & (-0.70, -0.47) & (-0.59, -0.58) & (-0.71, -0.46) & (-0.66, -0.58) & (-0.77, -0.46) \\ 
  \bf{physically active} &0.41 & ( 0.34,  0.48) & ( 0.41,  0.42) & ( 0.34,  0.49) & ( 0.41,  0.46) & ( 0.34,  0.53) \\ 
  \bf{time} & -0.14 & (-0.15, -0.14) & (-0.18, -0.14) & (-0.19, -0.14) & (-0.25, -0.14) & (-0.25, -0.14) \\ 
   \hline
  \end{tabular}
 \caption{Comparison of estimates of the fixed effects from a multilevel model and $95\%$ confidence intervals (ci) assuming MAR, and bounds for estimates and $95\%$ uncertainty intervals (ui) under two different MNAR assumptions. Age is rescaled to improve convergence.}
 \label{res.tab}
\end{table}

\section{Simulation study}
\label{Simulation}
\subsection{Simulation design}
\label{sim.design}
The purpose of the simulation study is to evaluate how well the bias correction performs in finite sample sizes and under various violations of the missing at random (MAR) assumption, in scenarios resembling the motivating example. For simplicity, we include only two covariates in addition to time. In all simulation designs, covariates are generated at baseline, with $x_1 \sim N(0,1)$ and $x_2 \sim \text{Be}(0.3)$. The third covariate is time since baseline, which—similar to the motivating example—takes values [0, 2, 4, 6] years for individuals without missing data.
Across all designs, we use $\bbeta = [2, 1, 0.5, 1]$ and $\bdelta = [2.5, -2, 2, -0.2]$. Designs 1 and 2 are nested (i.e., there is dependence between individuals within the same household), while Designs 3–4 are unnested (i.e., individuals within the same household are independent); see Table \ref{tab.rho} for exact parameter values.
 We use {\tt lme4} to estimate $\tilde \bbeta$, $\bdelta$ and all variance and covariance parameters in table \ref{tab.rho} except for $\rho_{\text{hh}}$, $\rho_{\text{ind}}$ and $\rho_{\text{err}}$, which must be pre-specified in order to estimate the bias due to MNAR. Estimates of the covariance matrices $\hat\Sigma$, $\hat\Omega$, and $\hat\Delta$ are then derived from these estimates ($\rho_{\text{hh}}$, $\rho_{\text{ind}}$ and $\rho_{\text{err}}$ are required for $\hat\Delta$). 
 We consider sample sizes of 500 (results in Appendix \ref{appRe}) and 2000 households, with two individuals per household, and perform 1000 replications. For each design, we fit two multilevel models (using {\tt lmer} and {\tt glmer}): one nested and one unnested. The exact model specifications are provided in Table \ref{tab.formulas}. All models are fitted using the optimizer bobyqa with a maximum of $10^6$ function evaluations, regardless of convergence issues.
Note that $\hat\Sigma$ may be slightly biased due to MNAR data. While this is not problematic for estimating $\tilde\bbeta$ (since the expected value remains unchanged), it may affect the derivation of standard errors and bias (as $\hat\Delta$ is derived using $\hat\Sigma$). However, we choose to ignore this potential issue, as we believe that any correction would not sufficiently improve the results to justify the additional computational cost.
  Finally, we use the estimates of the standard errors for $ \hat{\tilde \bbeta} $ (default estimates returned by {\tt lmer} using the incomplete data) as an approximation of the standards errors of $\hat \bbeta = \hat{\tilde \bbeta} - \widehat {bias(\tilde\bbeta)}$.

\begin{table}[h!]
\begin{tabular}{ c | c c c c c c c c c c c c c}
 &$\rho_{ac}$&$\rho_{bd}$&$\rho_{\text{hh}}$&$\rho_{\text{ind}}$&$\rho_{\text{err}}$& $\sigma_a$ &$\sigma_b$ &$\sigma_c$ &$\sigma_d$ &$\sigma_\nu$&$\sigma_\gamma$ &$\sigma_\xi$ &$\sigma_\epsilon$ \\
 \hline
 Design 1 & -0.2 & -0.2 & 0.3 & 0.3 & 0 &\multirow{2}{*}{2}&\multirow{2}{*}{0.5}&\multirow{2}{*}{0.5}&\multirow{2}{*}{0.5}&\multirow{2}{*}{0.3}&\multirow{2}{*}{0.5}&\multirow{2}{*}{1}&\multirow{2}{*}{2}\\ 
 Design 2& -0.2 & -0.2 & 0.3 & 0.3 & 0.3\\
 Design 3 & 0& -0.2& 0& 0.3& 0&\multirow{2}{*}{0}&\multirow{2}{*}{2}&\multirow{2}{*}{0}&\multirow{2}{*}{0.5}&\multirow{2}{*}{0}&\multirow{2}{*}{0.6}&\multirow{2}{*}{1}&\multirow{2}{*}{2}\\ 
 Design 4&0& -0.2& 0& 0& 0.5  
\end{tabular}
 \caption{Parameters that generate the covariance matrices of the data generating process. The standard deviations and correlations of the table refer the parameters of equations (\ref{Ex.out.eq}) - (\ref{Ex.sel.eq})  where for instance $\sigma_a$ is the standard deviation of $a$ and $\rho_{ab}$ is the correlation between $a$ and $b$. }
 \label{tab.rho}
\end{table}

\begin{table}[h!]
\begin{tabular}{ c  c c c c c c c c c c c c c}
& Fixed effects & \multicolumn{2}{c}{ Random effects} \\
&&Nested design&Unnested design\\
Outcome model& $y \sim x1+x2+time$& $+(1 + time | id/hhid)$&$+(1 + time | id)$\\
Dropout model&$z \sim x1+x2+time$&$+(0 + time | id/hhid)$&$+(0 + time | id)$
\end{tabular}
 \caption{Formulas used for estimation of parameters, where $id$ is individual identifier and $hhid$ is household identifier. The formula consists of the fixed effects and one of the two expressions of random effects.}
 \label{tab.formulas}
\end{table}

\subsection{Simulation results}
\label{sim.res}
Estimation of both the dropout and outcome models using a nested approach results in error messages in most replications across all designs (see Web Table 1). In contrast, the unnested model demonstrates greater stability: the outcome model is estimated without error messages in all replications, and the dropout model is estimated without error messages in over $80\%$ of the replications. Note that, we disregard all warnings related to poor convergence and proceed with the estimates provided by the functions.

\begin{figure}
\centering
\includegraphics[width=\textwidth]{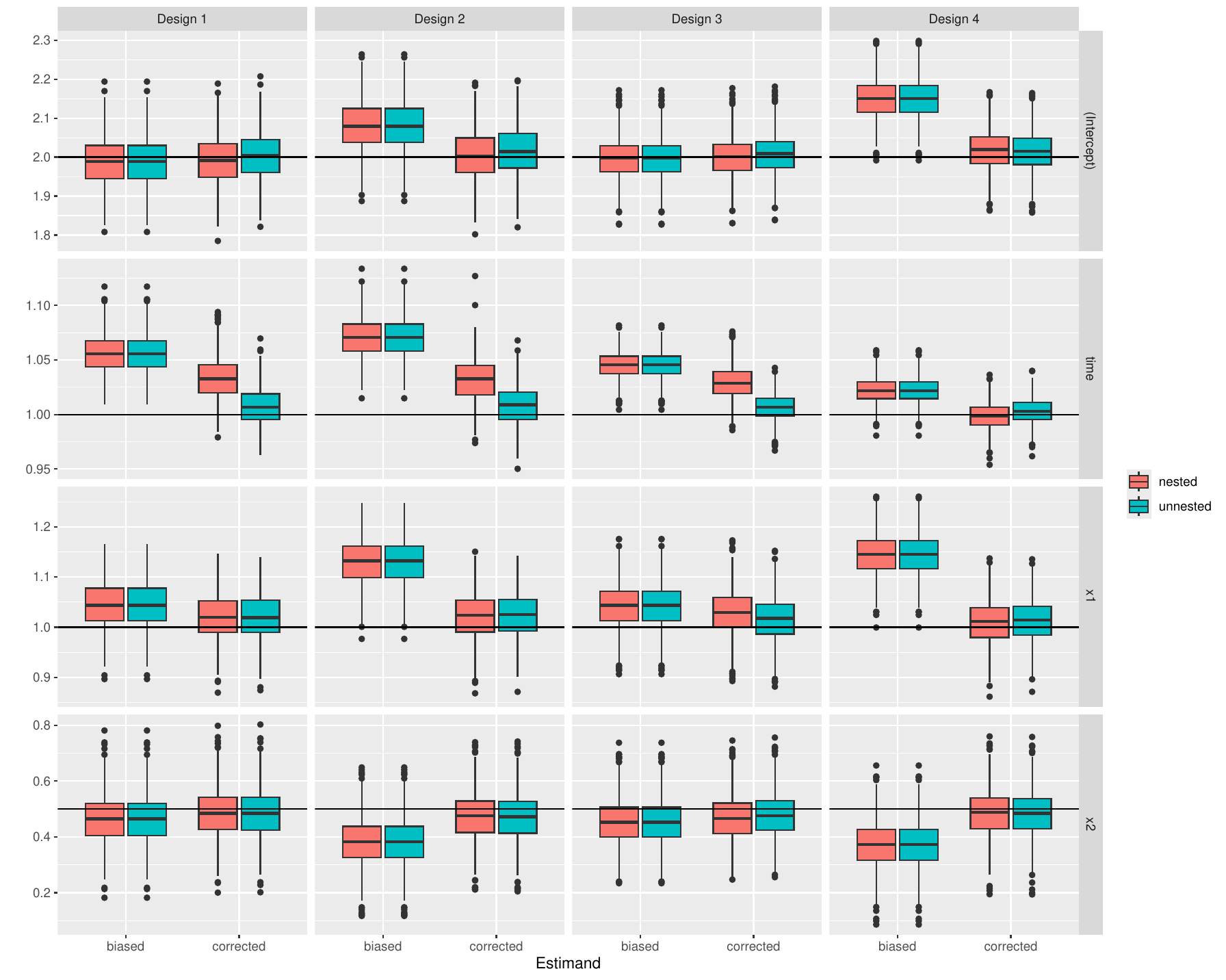}
\caption{Boxplot of the biased estimates and corrected biased estimates using the estimates from {\tt lmer} and {\tt glmer} along with the $\rho_{\text{hh}}$, $\rho_{\text{ind}}$ and $\rho_{\text{err}}$ that was used to generate data for each Design for the sample size with 2000 households.}
\label{boxplot2000}
\end{figure}

\begin{figure}
\centering
\includegraphics[width=\textwidth]{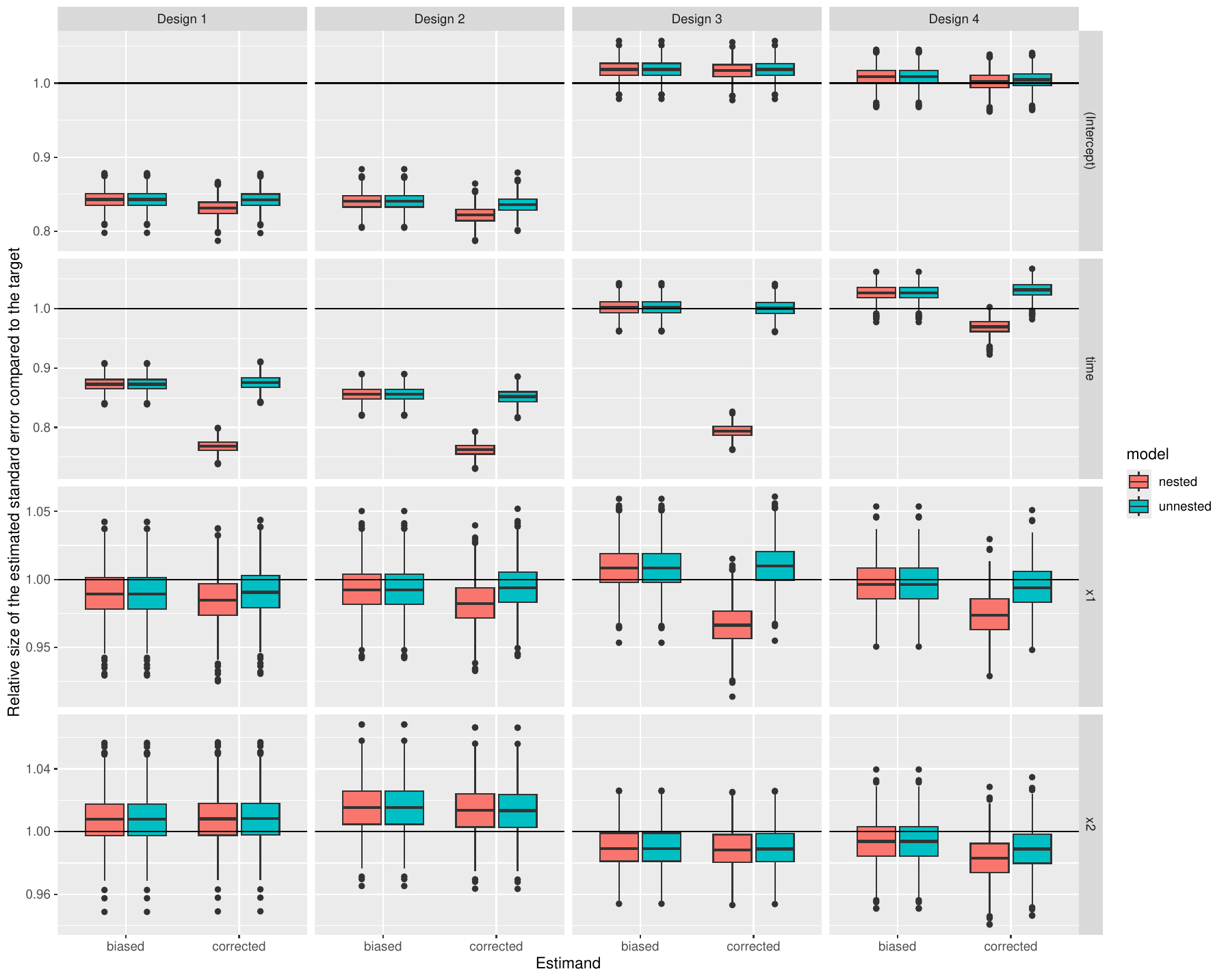}
\caption{Boxplot of the quotient between the estimated standard errors (default from lmer) and the Monte Carlo (mc) standard errors of the estimates of $\tilde \bbeta$ and $\bbeta$ (the bias correction was estimated using $\rho_{\text{hh}}$, $\rho_{\text{ind}}$ and $\rho_{\text{err}}$ from the data generating process). Note that, the estimated standard errors are the same for for the unnested and nested design up to 4 decimals, we also do not adjust the estimated standard errors to incorporate the bias correction when estimating $\bbeta$, therefore the difference in the quotient between within the same design and parameter is due to difference in the mc standard errors.}
\label{rel.se.2000}
\end{figure}

Using nested and unnested models yields nearly identical estimates of $\tilde \bbeta$ (i.e. default estimates returned by {\tt lmer} using the incomplete data) and corresponding standard errors. The main difference between the models lies in the bias correction: while the corrections are similar for the intercept, $x_1$, and $x_2$, they differ for the time variable. In this case, the unnested model performs better across all designs (see Figure \ref{boxplot2000}). In fact, the unnested model is nearly unbiased in all designs, including those generated under a nested structure (i.e. Design 1 and 2). 
As expected (see Section \ref{sim.design}), the estimated standard errors are not always unbiased (see Figure \ref{rel.se.2000}). This issue is most pronounced for the time variable in the nested design, where the estimated standard errors are only $70-80\%$ of the desired value for Designs 1–3, resulting in empirical coverage of approximately $40\%$ (see Table \ref{res.tab.2000} and \ref{res.tab.500}). For the other parameters and for Design 4, the standard errors appear to be adequate approximations. Consequently, the bias-corrected estimates achieve empirical coverage of at least $87\%$, and often between $90\%$ and $95\%$ (see Table \ref{res.tab.2000}). The results using a smaller sample size (500 households) are similar to these results, see Appendix \ref{appRe}.

\begin{table}[ht]
\centering
\rowcolors{2}{white}{gray!30}
\begin{tabular}{lllrrrrrrrrrrrrrrr}
 \rowcolor{white}
&\multicolumn{3}{c}{Design 1}&\multicolumn{3}{c}{Design 2}&\multicolumn{3}{c}{Design 3}&\multicolumn{3}{c}{Design 4}\\
\cmidrule(lr){2-4} \cmidrule(lr){5-7} \cmidrule(lr){8-10} \cmidrule(lr){11-13} 
 \rowcolor{white}
 estimand  & MSE & bias & cov & MSE & bias & cov & MSE & bias & cov & MSE & bias & cov  \\ 
  \hline
$\tilde \bbeta_{Int}$  &0.4 & -1.1 & 89.6 & 1.1 & 8.2 & 63.5 & 0.3 & -0.4 & 94.7 & 2.5 & 15.0 & 16.5 \\ 
  $\tilde \bbeta_{Int}$  & 0.4 & -1.1 & 89.6 & 1.1 & 8.2 & 63.5 & 0.3 & -0.4 & 94.7 & 2.5 & 15.0 & 16.5 \\ 
  $\bbeta_{Int}$  & 0.4 & -0.8 & 89.5 & 0.4 & 0.5 & 89.2 & 0.3 & -0.1 & 94.8 & 0.3 & 1.8 & 93.9 \\ 
  $ \bbeta_{Int}$  & 0.4 & 0.4 & 90.0 & 0.4 & 1.7 & 88.7 & 0.3 & 0.7 & 95.1 & 0.3 & 1.5 & 94.6 \\ 
$\tilde \bbeta_{time}$ &  0.3 & 5.6 & 6.6 & 0.5 & 7.1 & 0.6 & 0.2 & 4.5 & 3.7 & 0.1 & 2.2 & 57.3 \\ 
$\tilde \bbeta_{time}$ & 0.3 & 5.6 & 6.6 & 0.5 & 7.1 & 0.6 & 0.2 & 4.5 & 3.7 & 0.1 & 2.2 & 57.3 \\ 
$ \bbeta_{time}$ & 0.1 & 3.3 & 43.4 & 0.1 & 3.2 & 42.0 & 0.1 & 3.0 & 36.7 & 0.0 & -0.1 & 93.6 \\ 
$ \bbeta_{time}$ & 0.0 & 0.7 & 89.7 & 0.0 & 0.8 & 87.3 & 0.0 & 0.7 & 91.3 & 0.0 & 0.3 & 94.2 \\ 
  $\tilde \bbeta_{x1}$  &0.4 & 4.5 & 81.6 & 1.9 & 13.1 & 17.3 & 0.4 & 4.3 & 81.8 & 2.3 & 14.5 & 7.7 \\ 
  $\tilde \bbeta_{x1}$  &  0.4 & 4.5 & 81.6 & 1.9 & 13.1 & 17.3 & 0.4 & 4.3 & 81.8 & 2.3 & 14.5 & 7.7 \\ 
  $ \bbeta_{x1}$  &  0.3 & 2.1 & 92.0 & 0.3 & 2.2 & 92.0 & 0.3 & 3.0 & 88.3 & 0.2 & 1.0 & 94.1 \\ 
  $ \bbeta_{x1}$  &  0.2 & 2.1 & 92.2 & 0.3 & 2.4 & 91.3 & 0.2 & 1.7 & 93.4 & 0.2 & 1.3 & 94.1 \\ 
  $\tilde \bbeta_{x2}$  & 0.9 & -3.6 & 93.3 & 2.1 & -11.6 & 73.5 & 0.9 & -4.5 & 91.2 & 2.3 & -12.7 & 66.0 \\ 
  $\tilde \bbeta_{x2}$  & 0.9 & -3.6 & 93.3 & 2.1 & -11.6 & 73.5 & 0.9 & -4.5 & 91.2 & 2.3 & -12.7 & 66.0 \\ 
  $ \bbeta_{x2}$  & 0.8 & -1.5 & 95.1 & 0.8 & -2.6 & 94.4 & 0.8 & -3.3 & 92.5 & 0.7 & -1.4 & 94.8 \\ 
  $ \bbeta_{x2}$  & 0.8 & -1.6 & 94.7 & 0.8 & -2.8 & 94.3 & 0.7 & -2.2 & 93.1 & 0.7 & -1.6 & 94.7 \\ 
   \hline
\end{tabular}
\caption{Mean squrare error (MSE) as well as empirical bias and coverage of $95 \%$ confidence intervals for $\bbeta$, using the default estimates from lmer (i.e.the estimand $\tilde \bbeta$) and corrected for bias assuming correct values for the sensitivity parameters $\rho_{\text{hh}}$, $\rho_{\text{ind}}$ and $\rho_{\text{err}}$ (i.e. the estimand $\bbeta$). The whte rows have used a nested design in the model specification when estimating the multilevel model and the gray have an unnested design (see table \ref{tab.formulas}). The sample size is 2000 households. All values have been multiplied by $10^2$.} 
\label{res.tab.2000}
\end{table}

\section{Concluding remarks}
\label{Conclusion}
The aim with this paper was to develop analysis tools that allow for less restrictive assumptions than MAR when using multilevel models, as previous research indicate that MAR may not be fulfilled when investigating health and ageing. The proposed method can be used as a sensitivity analysis of the MAR assumption, enabling comparison of estimates under different missing data mechanisms. The simulation study seems to support that the bias correction works reasonably well in finite samples.
One limitation of the approach is that, although the theoretical framework accommodates large clusters, the computational burden becomes substantial when clusters contain many observations. The main challenge lies in numerically deriving the expected value of a doubly truncated multivariate distribution, which becomes computationally intensive in high dimensions. As a result, it is not feasible to implement a nested design with country as a hierarchical level, even though this would more accurately reflect the sampling scheme of the motivating example.

The sensitivity analysis was applied to a longitudinal study and suggests that the conclusions about the parameters of interest of a MAR analysis are similar under the milder assumptions that poor memory may be associated with higher dropout. However, some caution is warranted when interpreting the results, as the target population is somewhat ill-defined. Some individuals in the sample may have died before the final wave (wave 9), and selection into the study was not accounted for.

Finally, a future aim is to incorporate the method into the R package {\tt{ui}} (\citealp{ui}) which contain tools for similar sensitivity analyses for other types of regression models.

 \bibliography{referenser}

\begin{appendices}

\section{}\label{appTh}
\begin{lemma}
\begin{align*}
E\left(\y_i \left| \left[\z_i^{\text{obs}}=1 \atop \z_i^{mis}=0\right], \X_i \right. \right)&= \X_i \bbeta +\Delta_{i} \Omega_{i}^{-1}  E\left( \bu_i \left| \left[ \bu_i^{\text{obs}} > -{\X_i}^{\text{obs}} \delta \atop \bu_i^{mis} < -{\X_i}^{mis} \delta\right], \X_i \right. \right)
\end{align*}

\begin{proof}
Let $\w_i=\boeta_j-\Delta_{i} \Omega_{i}^{-1} \bu_i$. Then $\mbox{E}(\w_i)=\0$ and $\w_i$ is independent of $\bu_i$ since both $\w_i$ and $\bu_i$ are normally distributed and
\begin{align*}
\mbox{cov}(\w_i,\bu_i)=\mbox{cov}(\boeta_j,\bu_i)-\mbox{cov}(\Delta_{i} \Omega_{i}^{-1} \bu_i,\bu_i)=\Delta_{i}-\Delta_{i} \Omega_{i}^{-1} \Omega_{i}=\0.
\end{align*}
Therefore,
\begin{align*}
 &E\left( \boeta_i \left| \left[ \bu_i^{\text{obs}} > -{\X_i}^{\text{obs}} \delta \atop \bu_i^{mis} < -{\X_i}^{mis} \delta\right], \X_i \right. \right)= E\left(\w_i + \Delta_{i} \Omega_{i}^{-1} \bu_i \left| \left[ \bu_i^{\text{obs}} > -{\X_i}^{\text{obs}} \delta \atop \bu_i^{mis} < -{\X_i}^{mis} \delta\right], \X_i \right. \right)\\
&=E\left(\w_i  \left| \left[ \bu_i^{\text{obs}} > -{\X_i}^{\text{obs}} \delta \atop \bu_i^{mis} < -{\X_i}^{mis} \delta\right], \X_i \right. \right) + E\left( \Delta_{i} \Omega_{i}^{-1} \bu_i \left| \left[ \bu_i^{\text{obs}} > -{\X_i}^{\text{obs}} \delta \atop \bu_i^{mis} < -{\X_i}^{mis} \delta\right], \X_i \right. \right)\\
&=\0 + \Delta_{i} \Omega_{i}^{-1}  E\left( \bu_i \left| \left[ \bu_i^{\text{obs}} > -{\X_i}^{\text{obs}} \delta \atop \bu_i^{mis} < -{\X_i}^{mis} \delta\right], \X_i \right. \right)
\end{align*}

\end{proof}
\label{proof.exp}
\end{lemma}

\section{}\label{appRe}

In this appendix, we present the frequency of different error codes observed in the simulations (see Tables~\ref{err.tab.2000} and~\ref{err.tab.500}), along with simulation results for a sample size of 500 households. These include the corresponding figures and tables as in Section~4.2. The results for 500 households are similar to those obtained with a sample size of 2000 households.

\begin{table}[ht]
\centering
\begin{tabular}{rrrrrrrrrrrrr}
  \hline
&&\multicolumn{2}{c}{Design 1}&\multicolumn{2}{c}{Design 2}&\multicolumn{2}{c}{Design 3}&\multicolumn{2}{c}{Design 4}\\
\cmidrule(lr){3-4} \cmidrule(lr){5-6} \cmidrule(lr){7-8} \cmidrule(lr){9-10} 
&&glmer& lmer & glmer & lmer & glmer & lmer & glmer & lmer  \\ 
  \hline
\multirow{10}{*}{%
        \rotatebox[origin=c]{90}{nested formula}}&  Message 1 & 132 & 459 & 134 & 479 & 135 & 441 & 145 & 453 \\ 
 & Message 2 & 629 & 248 & 645 & 259 & 604 & 345 & 619 & 332 \\ 
 & Message 3 & 365 & 390 & 348 & 372 & 385 & 350 & 376 & 376 \\ 
 & Message 4 & 364 & 390 & 348 & 372 & 385 & 350 & 376 & 376 \\ 
 & Message 5 & 1 & 22 & 0 & 21 & 0 & 113 & 0 & 89 \\ 
 & Message 6 & 1 & 0 & 0 & 0 & 0 & 0 & 0 & 0 \\ 
  &total = 0  & 0 & 68 & 1 & 73 & 0 & 32 & 0 & 34 \\ 
  &total = 1 & 511 & 355 & 527 & 351 & 495 & 337 & 486 & 306 \\ 
  &total = 2 & 486 & 577 & 468 & 576 & 501 & 631 & 512 & 660 \\ \smallskip\smallskip
  & total = 3 & 3 & 0 & 4 & 0 & 4 & 0 & 2 & 0 \\ 
\multirow{10}{*}{%
        \rotatebox[origin=c]{90}{unnested formula}} & Message 1 & 162 & 0 & 168 & 0 & 176 & 0 & 169 & 0 \\ 
 & Message 2 & 147 & 0 & 158 & 0 & 163 & 0 & 161 & 0 \\ 
 & Message 3 & 6 & 0 & 4 & 0 & 3 & 0 & 1 & 0 \\ 
 & Message 4 & 6 & 0 & 4 & 0 & 3 & 0 & 1 & 0 \\ 
 & Message 5 & 0 & 0 & 0 & 0 & 0 & 0 & 0 & 0 \\ 
 & Message 6 & 0 & 0 & 0 & 0 & 0 & 0 & 0 & 0 \\ 
  &total = 0 & 832 & 1000 & 828 & 1000 & 821 & 1000 & 830 & 1000 \\ 
  &total = 1 & 16 & 0 & 14 & 0 & 13 & 0 & 13 & 0 \\ 
  &total = 2 & 151 & 0 & 154 & 0 & 166 & 0 & 152 & 0 \\ 
  &total = 3 & 1 & 0 & 4 & 0 & 0 & 0 & 5 & 0 \\ 
     \hline
\multicolumn{10}{l}{\footnotesize Message 1 = "Model failed to converge with max$|$grad$|$ = ... "}\\ 
\multicolumn{10}{l}{\footnotesize Message 2 = "Model is nearly unidentifiable: large eigenvalue ratio
 - Rescale variables?"} \\ 
 \multicolumn{10}{l}{\footnotesize Message 3 = "Model failed to converge: degenerate  Hessian with ? negative eigenvalues"} \\ 
  \multicolumn{10}{l}{\footnotesize Message 4 = "unable to evaluate scaled gradient"} \\ 
  \multicolumn{10}{l}{\footnotesize Message 5 = "boundary (singular) fit: see help('isSingular')"} \\ 
  \multicolumn{10}{l}{\footnotesize Message 6 = "Hessian is numerically singular: parameters are not uniquely determined"} \\ 
   \hline
\end{tabular}
\caption{Frequency of different error messages from glmer (missing data regression) and lmer (outcome regression) (all error messages were ignored) for the different simulation designs using sample size with 2000 households} 
\label{err.tab.2000}
\end{table}

\begin{table}[ht]
\centering
\begin{tabular}{rrrrrrrrrrrrr}
  \hline
&&\multicolumn{2}{c}{Design 1}&\multicolumn{2}{c}{Design 2}&\multicolumn{2}{c}{Design 3}&\multicolumn{2}{c}{Design 4}\\
\cmidrule(lr){3-4} \cmidrule(lr){5-6} \cmidrule(lr){7-8} \cmidrule(lr){9-10} 
&&glmer& lmer & glmer & lmer & glmer & lmer & glmer & lmer  \\ 
  \hline
\multirow{10}{*}{%
        \rotatebox[origin=c]{90}{nested formula}}&Message 1 &93 & 5 & 83 & 0 & 83 & 7 & 85 & 9 \\ 
&  Message 2 &548 & 455 & 535 & 490 & 559 & 452 & 573 & 452 \\ 
 & Message 3 & 425 & 505 & 432 & 475 & 408 & 495 & 402 & 491 \\ 
 & Message 4 & 424 & 505 & 429 & 475 & 407 & 495 & 402 & 491 \\ 
 & Message 5 & 1 & 33 & 1 & 29 & 0 & 53 & 1 & 57 \\ 
 & Message 6 &2 & 0 & 4 & 0 & 1 & 0 & 0 & 0 \\ 
 & total = 0 & 0 & 7 & 0 & 6 & 0 & 0 & 0 & 0 \\ 
 & total = 1 & 509 & 483 & 518 & 519 & 544 & 498 & 539 & 500 \\ 
 & total = 2 & 489 & 510 & 480 & 475 & 454 & 502 & 459 & 500 \\  \smallskip\smallskip
 & total = 3 &2 & 0 & 2 & 0 & 2 & 0 & 2 & 0 \\ 
\multirow{10}{*}{%
        \rotatebox[origin=c]{90}{unnested formula}}&  Message 1 & 105 & 0 & 119 & 0 & 91 & 0 & 112 & 0 \\ 
&  Message 2 & 70 & 0 & 80 & 0 & 65 & 0 & 76 & 0 \\ 
&  Message 3 & 2 & 0 & 3 & 0 & 3 & 0 & 2 & 0 \\ 
 & Message 4 &2 & 0 & 3 & 0 & 3 & 0 & 2 & 0 \\ 
&  Message 5 &0 & 0 & 0 & 0 & 0 & 0 & 0 & 0 \\ 
&  Message 6 & 0 & 0 & 0 & 0 & 0 & 0 & 0 & 0 \\ 
&  total = 0  & 893 & 1000 & 878 & 1000 & 906 & 1000 & 886 & 1000 \\ 
&  total = 1 &  35 & 0 & 39 & 0 & 26 & 0 & 36 & 0 \\ 
&  total = 2 &72 & 0 & 83 & 0 & 68 & 0 & 78 & 0 \\ 
&  total = 3 &  0 & 0 & 0 & 0 & 0 & 0 & 0 & 0 \\
   \hline

\multicolumn{12}{l}{\footnotesize Message 1 = "Model failed to converge with max$|$grad$|$ = ... "}\\ 
\multicolumn{12}{l}{\footnotesize Message 2 = "Model is nearly unidentifiable: large eigenvalue ratio
 - Rescale variables?"} \\ 
 \multicolumn{12}{l}{\footnotesize Message 3 = "Model failed to converge: degenerate  Hessian with ? negative eigenvalues"} \\ 
  \multicolumn{12}{l}{\footnotesize Message 4 = "unable to evaluate scaled gradient"} \\ 
  \multicolumn{12}{l}{\footnotesize Message 5 = "boundary (singular) fit: see help('isSingular')"} \\ 
  \multicolumn{12}{l}{\footnotesize Message 6 = "Hessian is numerically singular: parameters are not uniquely determined"} \\ 
   \hline
\end{tabular}
\caption{Frequency of different error messages from glmer (missing data regression) and lmer (outcome regression) (all error messages were ignored) for the different simulation designs using sample size with 500 households} 
\label{err.tab.500}
\end{table}

\begin{figure}
\centering
\includegraphics[width=\textwidth]{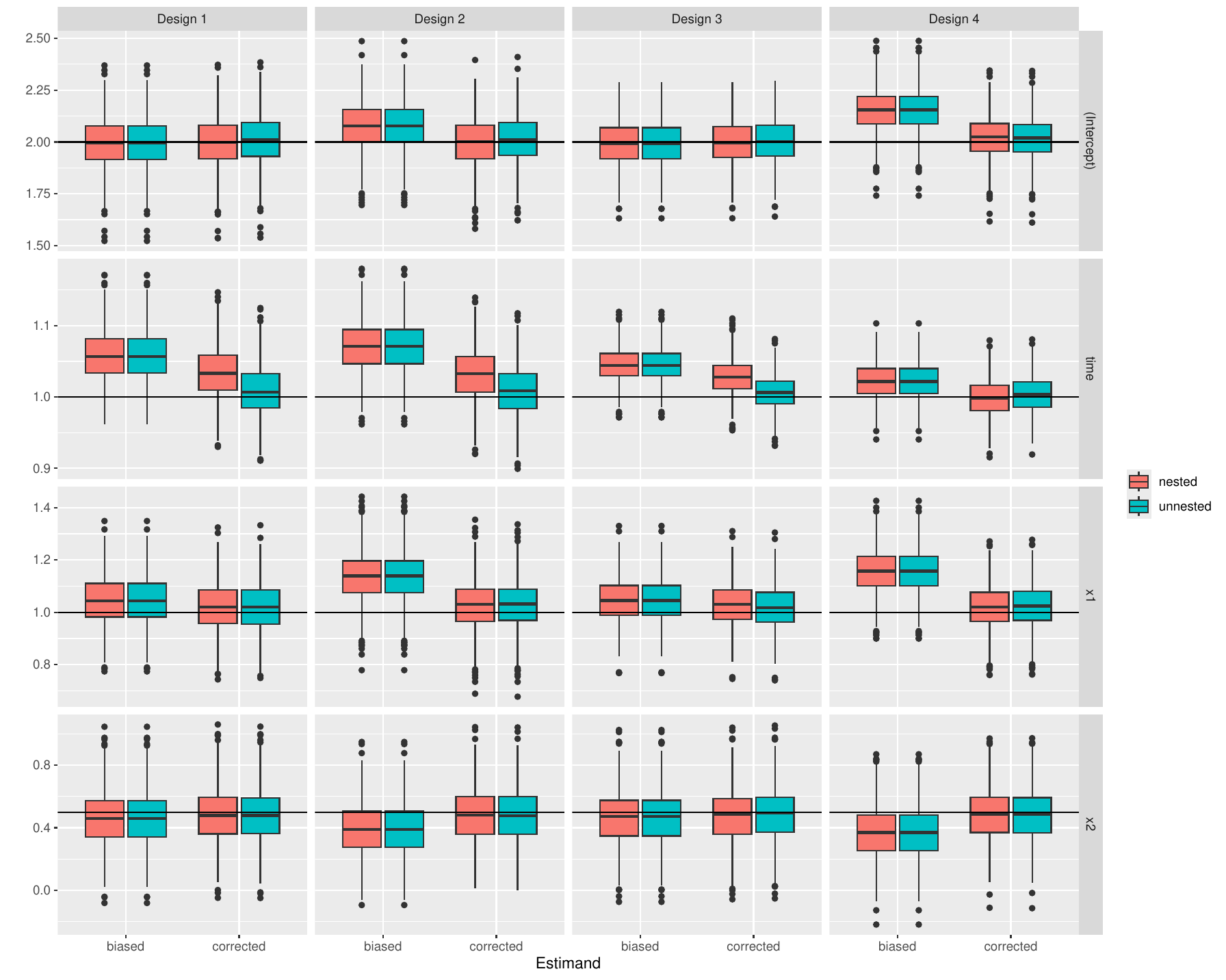}
\caption{Boxplot of the biased estimates and corrected biased estimates using the estimates from {\tt lmer} and {\tt glmer} along with the $\rho_{\text{hh}}$, $\rho_{\text{ind}}$ and $\rho_{\text{err}}$ that was used to generate data for each Design for the sample size with 500 households.}
\label{boxplot500}
\end{figure}

\begin{figure}
\centering
\includegraphics[width=\textwidth]{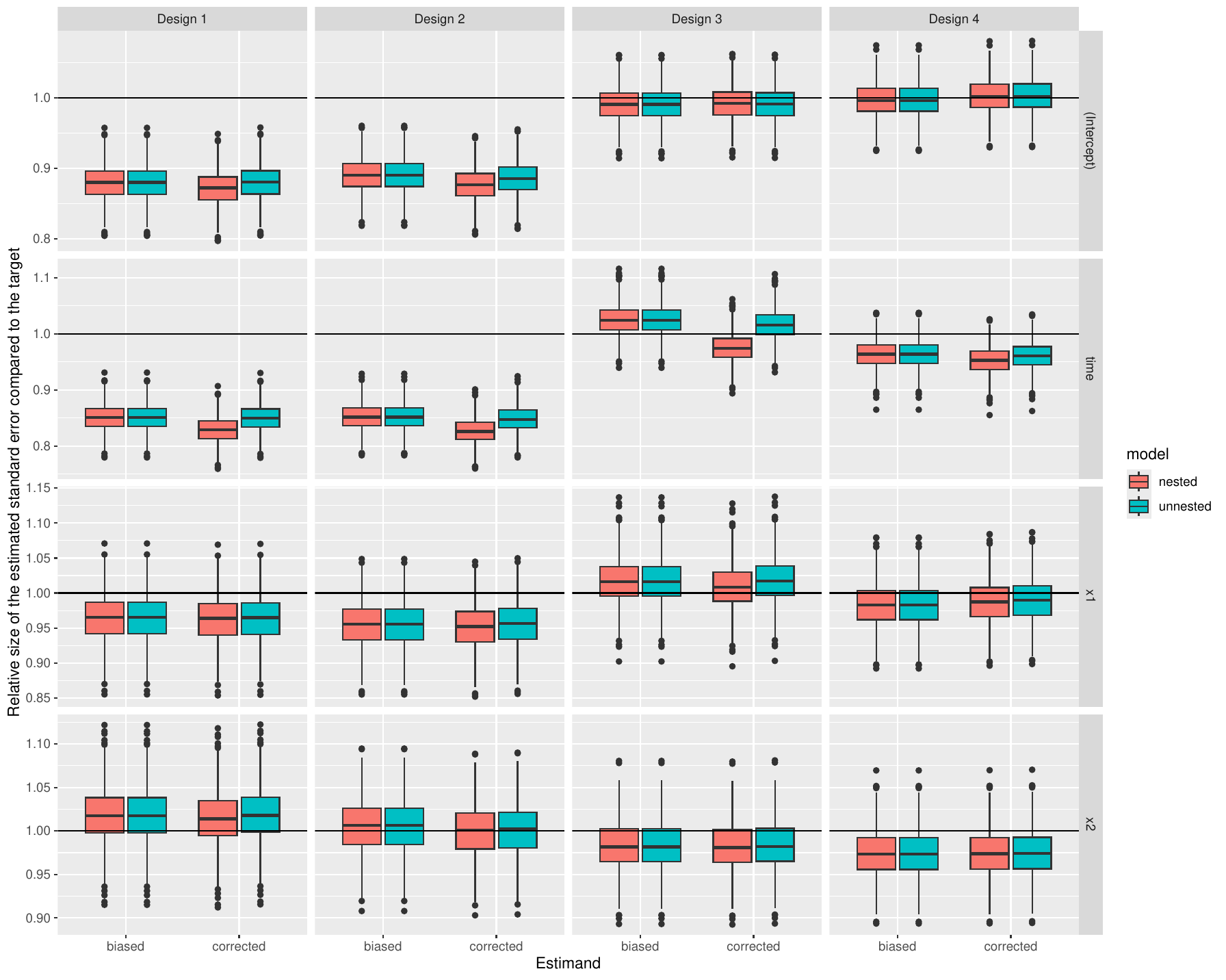}
\caption{Boxplot of the quotient between the estimated standard errors (default from lmer) and the Monte Carlo (mc) standard errors of the estimates of $\tilde \bbeta$ and $\bbeta$ (the bias correction was estimated using $\rho_{\text{hh}}$, $\rho_{\text{ind}}$ and $\rho_{\text{err}}$ from the data generating process) for the sample size with 500 households. Note that, the estimated standard errors are the same for for the unnested and nested design up to 4 decimals, we also do not adjust the estimated standard errors to incorporate the bias correction when estimating $\bbeta$, therefore the difference in the quotient between within the same design and parameter is due to difference in the mc standard errors.}
\label{rel.se.500}
\end{figure}

\begin{table}[ht]
\centering
\rowcolors{2}{white}{gray!30}
\begin{tabular}{lllrrrrrrrrrrrrrrr}
 \rowcolor{white}
&\multicolumn{3}{c}{Design 1}&\multicolumn{3}{c}{Design 2}&\multicolumn{3}{c}{Design 3}&\multicolumn{3}{c}{Design 4}\\
\cmidrule(lr){2-4} \cmidrule(lr){5-7} \cmidrule(lr){8-10} \cmidrule(lr){11-13}
 \rowcolor{white}
 estimand  & MSE & bias & cov & MSE & bias & cov & MSE & bias & cov & MSE & bias & cov \\ 
  \hline
$\tilde \bbeta_{Int}$  &1.5 & -0.4 & 91.2 & 2.0 & 7.7 & 86.0 & 1.1 & -0.7 & 94.9 & 3.4 & 15.3 & 68.8 \\ 
$\tilde \bbeta_{Int}$  & 1.5 & -0.4 & 91.2 & 2.0 & 7.7 & 86.0 & 1.1 & -0.7 & 94.9 & 3.4 & 15.3 & 68.8 \\ 
  $ \bbeta_{Int}$  & 1.5 & -0.2 & 90.7 & 1.5 & 0.0 & 90.7 & 1.1 & -0.4 & 95.1 & 1.1 & 2.2 & 95.0 \\ 
  $\bbeta_{Int}$  & 1.5 & 1.1 & 90.6 & 1.5 & 1.2 & 91.1 & 1.1 & 0.4 & 94.9 & 1.1 & 1.9 & 94.9 \\ 
$\tilde \bbeta_{time}$ & 0.4 & 5.7 & 52.1 & 0.6 & 7.0 & 35.8 & 0.3 & 4.5 & 55.3 & 0.1 & 2.2 & 84.4 \\ 
$\tilde \bbeta_{time}$ &  0.4 & 5.7 & 52.1 & 0.6 & 7.0 & 35.8 & 0.3 & 4.5 & 55.3 & 0.1 & 2.2 & 84.4 \\ 
$\bbeta_{time}$ & 0.2 & 3.4 & 73.8 & 0.2 & 3.1 & 75.3 & 0.1 & 2.8 & 78.5 & 0.1 & -0.1 & 94.0 \\ 
$\bbeta_{time}$ & 0.1 & 0.8 & 89.2 & 0.1 & 0.8 & 89.5 & 0.1 & 0.7 & 94.6 & 0.1 & 0.3 & 93.6 \\ 
  $\tilde \bbeta_{x1}$  &1.1 & 4.4 & 91.6 & 2.8 & 13.7 & 64.9 & 0.9 & 4.6 & 91.8 & 3.1 & 15.4 & 54.9 \\ 
  $\tilde \bbeta_{x1}$  & 1.1 & 4.4 & 91.6 & 2.8 & 13.7 & 64.9 & 0.9 & 4.6 & 91.8 & 3.1 & 15.4 & 54.9 \\ 
  $ \bbeta_{x1}$  &0.9 & 2.0 & 94.1 & 1.0 & 2.8 & 92.1 & 0.8 & 3.0 & 93.4 & 0.8 & 1.8 & 94.0 \\ 
  $ \bbeta_{x1}$  & 0.9 & 2.0 & 94.0 & 1.0 & 3.0 & 92.6 & 0.7 & 2.0 & 94.9 & 0.8 & 2.2 & 94.2 \\
  $\tilde \bbeta_{x2}$  & 3.1 & -4.0 & 94.7 & 4.2 & -11.0 & 90.4 & 3.0 & -3.5 & 93.7 & 4.6 & -13.0 & 87.5 \\ 
  $\tilde \bbeta_{x2}$  &3.1 & -4.0 & 94.7 & 4.2 & -11.0 & 90.4 & 3.0 & -3.5 & 93.7 & 4.6 & -13.0 & 87.5 \\ 
  $ \bbeta_{x2}$  & 3.0 & -1.9 & 95.4 & 3.1 & -1.9 & 95.0 & 2.9 & -2.2 & 94.1 & 2.9 & -1.6 & 93.5 \\ 
  $\bbeta_{x2}$  & 3.0 & -2.0 & 95.2 & 3.1 & -2.1 & 94.8 & 2.8 & -1.2 & 94.0 & 2.9 & -1.9 & 93.5 \\ 
   \hline
\end{tabular}
\caption{Mean squrare error (MSE) as well as empirical bias and coverage of $95 \%$ confidence intervals for $\bbeta$, using the default estimates from lmer (i.e.the estimand $\tilde \bbeta$) and corrected for bias assuming correct values for the sensitivity parameters $\rho_{\text{hh}}$, $\rho_{\text{ind}}$ and $\rho_{\text{err}}$ (i.e. the estimand $\bbeta$). The white rows have used a nested design in the model specification when estimating the multilevel model and the gray have an unnested design (see table 4). The sample size is 500 households. All values have been multiplied by $10^2$.} 
\label{res.tab.500}
\end{table}

\end{appendices}

\end{document}